%% file: main.tex
\documentclass[10pt]{article} 
\usepackage[top=3cm,bottom=3cm,left=2.5cm,right=3cm]{geometry}

\usepackage{cite}

\usepackage{epsfig}
\usepackage{graphicx}

\usepackage{hyperref}
\hypersetup{colorlinks,citecolor=black,filecolor=black,linkcolor=black,
urlcolor=black,bookmarksnumbered}

\usepackage{graphicx}
\usepackage{dcolumn}
\usepackage{bm}
\usepackage{amsmath,amsfonts,amssymb,amsthm,amscd,bbm,mathrsfs}
\usepackage{listings}
\usepackage{colortbl}
\usepackage{booktabs}

\providecommand{\norm}[1]{\lVert #1\rVert }

\theoremstyle{definition}

\usepackage{bbm}
\numberwithin{equation}{section}

\title{
A Dual Number Approach for Numerical Calculation of derivatives and its use in the Spherical 4R Mechanism}
\author{
$^1$F. Pe\~nu\~nuri\footnote{francisco.pa@uady.mx} ,~ 
$^1$R. Pe\'on-Escalante\footnote{rpeon@uady.mx} ,~$^1$C. Villanueva\footnote{cesar.villanueva@uady.mx} ,~
$^2$Carlos A. Cruz-Villar\footnote{cacruz@cinvestav.mx}\\
{\footnotesize \textit{$^1$Facultad de Ingenier\'ia, Universidad Aut\'onoma de Yucat\'an, A.P. 150, Cordemex, M\'erida, Yucat\'an, M\'exico.}}\\
{\footnotesize \textit{$^2$Cinvestav-IPN, Departamento de Ingenier\'ia El\'ectrica, Av. IPN 2508, A. P. 14-740, 07300, M\'exico D.F., M\'exico.}}
}

\date{\today}

\begin{document}
\maketitle
\begin{abstract}
\input{abstract.tex}

\end{abstract}

\input{introduction.tex}
\input{sec2.tex}

\input{sec3.tex}
\input{sec4.tex}
\input{conclusions.tex}
\bibliography{ref}

\end{document}

%% file: abstract.tex
This paper proposes a methodology to calculate both the first and second 
derivatives of a vector function of one variable in a single computation step. 
The method is based on the nested application of the dual number approach for first order derivatives.
 It has been implemented in  Fortran language, a module which contains the  dual version of 
elementary functions as well as more complex functions, which are common in the 
field of rotational kinematics. Since we have three quantities of interest, namely the function itself and its first and second derivative, our basic numerical entity has three elements. Then, for a given vector function  
$f:\mathbb{R}\to \mathbb{R}^m$, its dual version will have the form 
$\tilde{f}:\mathbb{R}^3\to \mathbb{R}^{3m}$.  
 As a study case, the proposed methodology is used 
to calculate the velocity and acceleration of a point moving on the coupler-point 
curve  generated  by a spherical four-bar mechanism.

%% file: introduction.tex
\section{Introduction}
The calculation of  velocity and acceleration is often needed in the fields of 
physics and engineering. In some cases, there is no possibility to obtain them 
analytically. In other cases, the analytical result is quite complicated to deal 
with; in both situations a numerical treatment is desirable. Regarding the field 
of mechanisms, the calculation of the first and second derivative allows the 
designer to include the velocity and acceleration in the synthesis of balanced 
mechanisms, dwell mechanisms, etc., \cite{Giacinto1981, Russella2005,Sandgren1985,
Umit2007,Mohan2009,Zobairi2003, Jiang2007}.

Usually, the process of calculating a derivative is not difficult. However, 
for the case of a spherical mechanism, obtaining first and second derivatives of 
the position vector for the coupler point is not simple. Even when such derivatives 
can be explicitly obtained, the resulting expressions could be of great complexity 
and useless for practical purposes. An alternative solution is to numerically 
calculate such derivatives. Nevertheless, traditional methods for calculating 
numerical derivatives (finite-differences) are subject to both truncation and 
subtractive cancellation errors, not to mention that they are not efficient enough to be used in the optimum synthesis of mechanisms. A different approach that is not subject to the above mentioned errors is automatic differentiation (AD) \cite{Neidinger2010}, an algorithmic approach to obtain derivatives for functions which are implemented in computer programs. AD can be implemented in several ways \cite{Griewank1989}, but the use of dual numbers is specially suited for that, since the chain rule can be implemented almost directly. 

 
Analogous to the definition of a complex number $z=a+i \,b$ with $a$ and  $b$ 
being real numbers and $i^2=-1$, a dual number is defined as $\hat{r}= a+ 
\epsilon \,b$ where $a$ and $b$ are real numbers but $\epsilon^2=0$. Although 
the first applications of the dual number theory to the development of mechanical 
engineering date back to early XX century \cite{Study1901}, and algebra of dual numbers was developed in the late XIX century \cite{Clif1873}, 
applications of dual numbers to numerical calculation of derivatives are relatively 
recent \cite{Leuck1999,Piponi2004,Jefrey2011}. Since then, a big amount of work 
regarding applications of dual numbers has been made. They are used to describe 
finite displacements of rigid and deformable bodies, for the  analytical treatment 
in kinematic and dynamics of spatial mechanisms, for the study of the kinematics, 
dynamics and calibration of open-chain robot manipulators, for the study of 
computer graphics, etc. In references  \cite{Jefrey2011, Cheng1994, Ettore2008} 
there is some bibliography about such studies.

Due to the great applicability of the dual numbers, it is important to develop 
algorithms implementing their algebra, which has been done in \cite{Cheng1994}. 
It is worthwhile to mention reference \cite{Ettore2008} where some linear dual 
algebra algorithms are presented.  Regarding applications of dual numbers to 
compute the numerical derivative of functions, we can cite 
\cite{Leuck1999,Piponi2004,Jefrey2011}. In  \cite{Leuck1999},  
dual numbers are used to calculate first order derivatives and in 
\cite{Piponi2004} and \cite{Jefrey2011}, they are used to calculate 
second order derivatives by using the operator overloading method.

The aim of this paper is twofold. First, it develops a methodology based on 
dual numbers where first and second order derivatives are 
obtained in a single computation step.  Second, it obtains the velocity and 
acceleration of a point moving on the coupler-point curve generated by a spherical 
four-bar mechanism. 

Once the methodology of developing elementary functions in its dual version including 
its second derivative has been presented, it is straightforward to develop more 
sophisticated functions such as rotations and numerical solutions to equations. Moreover, as the use of functions in programming languages as C, C++ and Fortran is quite intuitive 
and easy to codify, we have created a Fortran module  where  
some  functions in their dual version, as well as common functions to the field of 
rotational kinematics are provided. Such a module along with an example of its usage 
can be downloaded via internet 
at \url{http://www.meca.cinvestav.mx/personal/cacruz/archivos-ccv/}.

The rest of the paper is organized as follows. Section \ref{sec2} presents the 
process of calculating  numerical derivatives using  dual  numbers. 
Section \ref{sec3} presents the implementation in Fortran language of 
the dual  versions of scalar elementary functions as well as more 
complicated expressions, such as vector and matrix functions. In section \ref{sec4} 
we show how the velocity and acceleration of a point moving on the coupler-point 
curve generated by a spherical four-bar mechanism are obtained. Finally, section 
\ref{sectionconclu} presents the conclusions.

%% file: sec2.tex
\section{ Dual  numbers and derivatives}\label{sec2}
A dual number $\hat{r}$ is a number of the form 
\begin{equation}
\hat{r}= a +\epsilon \,b,
\end{equation}
where $a$ (the real part) and $b$ (the dual part) are real numbers and $\epsilon^2=0$. As in the case of complex numbers, there is an isomorphism\footnote{Under the ordinary addition and scalar multiplication---multiplication by a real number, the transformation $\mathbf{T}(a+b\,\epsilon)=\{a,b\}$ is linear and bijective.} 
between dual numbers and the real vector space $\mathbb{R}^2$. So, a dual number $\hat{r}$ can be defined as ordered pairs
\begin{equation}\label{xDual}
\hat{r}=\{a,b\}.
\end{equation} 
The algebraic rules for dual numbers can be found elsewhere in the literature, see for example  \cite{Clif1873, Brodsky1999,Ettore2008}.
Below we present how the dual numbers are used to calculate derivatives of a function.

\subsection{First order derivative} \label{subsec21}
Let us consider the Taylor series expansion (\ref{TayS}) of a function 
$f:\mathbb{R}\to \mathbb{R}$ about the point $x$, where $h.o.t.$ stands 
for higher order terms: 
\begin{equation} \label{TayS}
f(x+h)=f(x) + f'(x) h + \frac{f''(x)}{2} h^2 + \dots + h.o.t.
\end{equation}

Now, let us consider the dual number $\hat{x}=x+\epsilon$, i.e., a dual number 
where the coefficient of the nilpotent $\epsilon$ is equal to one. Substituting 
$\hat{x}$ in (\ref{TayS}), we obtain $ \hat{f}(\hat{x}) = f(x+\epsilon) = f(x) + f'(x) \epsilon. $

Instead of computing over the reals, we compute over the dual numbers and come up with a 
dual function $\hat{f} = f(x+\epsilon)$ whose real term is the original 
function $f(x)$ and the coefficient of $\epsilon$ is its derivative $f'(x)$. In the notation of (\ref{xDual}) we may write
\begin{equation} \label{dfv}
\hat{f}(\hat{x}) = \left\lbrace f(x),f'(x) \right\rbrace.
\end{equation} 

Let us exemplify the procedure for computing the dual function, by using the 
sinusoidal function. Let $f(x)=\sin (x)$ be the function for which the dual 
function has to be obtained and let $\hat{g}(\hat{x})=
\left\lbrace g_1(x),g_2(x) \right\rbrace$ be a dual function where $g_1(x)=g(x)$ 
and $g_2(x)=g'(x)$.  Then, from  (\ref{dfv}) we obtain 
\begin{equation} \label{sind}
\widehat{\sin}  (\hat{g}(\hat{x})) =\left\lbrace \sin \left(g_1(x)\right),
\cos \left(g_1(x)\right)g_2(x)\right\rbrace.
\end{equation}


\subsection{Second order derivative}
The second order derivative can be obtained by applying (\ref{dfv}) to $f'(x)$, that is,
stating the dual version of the function $f'(x)$ as
\begin{equation}
\hat{f} '(x)=  \left\lbrace f'(x),f''(x) \right\rbrace.
\end{equation}

The relevant information can be stored in a vector of three components.
 Such a vector  will have the information of the function, its first and its second derivative, 
so we could speak of an extended dual function (to differentiate it from the common dual function which has only two components).  We will use the notation 
\begin{equation}
\tilde{f}(x)=\left\lbrace f(x),f'(x),f''(x) \right\rbrace
\end{equation}
to represent the extended dual version of the original function $f$.
The identity function in its extended dual version $\tilde{x}$, has three components 
and is given by
\begin{equation} \label{idenD2}
\tilde{x}=\left\lbrace x,1,0 \right\rbrace.
\end{equation}

Let us exemplify the proposed methodology using the sinusoidal function. 
Let 
$$\tilde{g}(\tilde{x}) = \left\lbrace g_1(x),g_2(x),g_3(x) \right\rbrace$$
 be an 
extended dual  function, where $g_1(x)=g(x)$,  $g_2(x)=g'(x)$, $g_3(x)=g''(x)$. 
The sinusoidal function in its extended dual version is given by
\begin{align} \label{sineD2}
\widetilde{sin}(\tilde{g})  = & \{\sin(g_1),\; \cos(g_1)\,g_2,\; -\sin(g_1)\,g_2^2 \,+ \cos(g_1)\,g_3 \}
\end{align}
where the arguments of the functions are not written to simplify notation. Notice that no matter how complicated the function $g$, Eq. (\ref{sineD2}) ensures that the chain rule will be successfully applied. Thus by writing all the functions in their extended dual version the derivatives will be obtained without the need of traditional methods of finite-differences.

%% file: sec3.tex
\section{Fortran implementation of the extended dual functions}\label{sec3}

The extended dual functions are defined as arrays of $3m$ components and 
since we are interested in functions $f:\mathbb{R}\to \mathbb{R}^m$, a 
function $g:\mathbb{R}^3\to \mathbb{R}^{3m}$ is required to write the 
extended dual version of the function $f$. For example, in the case of 
the sinusoidal function and considering Fortran programming language, 
the code results as follows:

\begin{verbatim}
module sphmodual
	contains
	
function sindual(x) result(f_result)
	implicit none
	double precision :: f_result(3)
	double precision, intent(in) ::  x(3)
	f_result=[sin(x(1)),cos(x(1))*x(2),-sin(x(1))*x(2)**2 +	cos(x(1))*x(3)]
	return
end function sindual
	
end module sphmodual
\end{verbatim}
For coding purposes we will use {\tt fdual} instead of  $\tilde{f}$, so in the above code {\tt sindual} means $\widetilde{\sin}$.
 A simple program calculating the function $f(x)=\sin(\sin(x))$ for $x=1.1$ in 
 its extended dual version is:
 
 \begin{verbatim}
program sind
	use sphmodual
	implicit none
	double precision :: angd(3), res(3)
	angd=[1.1d0,1d0,0d0]
	res = sindual(sindual(angd))
	print*,res(1),res(2),res(3)
end program sind 
 \end{verbatim}
 
 After compiling and executing the above program, the result is:
 \begin{verbatim}
 0.777831   0.285073    -0.720138
 \end{verbatim}
The first component of the output vector corresponds to $f$, the second one to $f'$ and the third one to $f''$, all of them evaluated at the argument $x = 1.1$. The implementation of all of the other elementary functions is straightforward from this example. 


\subsection{Extended dual version of more complicated objects}
It has been shown how elementary functions can be \textit{dualized}, 
but there are more complicated objects such as, for example, the vector product, 
whose extended dual version is required for applications to rotational kinematics. 
One approach to handle vector functions is to work by components. Let us exemplify 
for the cross product.  The $i$-th component of the cross product of two vectors 
$\mathbf{a}$ and $\mathbf{b}$ is given by $(\mathbf{a} \times \mathbf{b})_i = \varepsilon_{ijk} a_j b_k $,
where $a_n$ and $b_n$ 
are the $n$-th component of vectors $\mathbf{a}$ and $\mathbf{b}$, respectively, and 
$\varepsilon_{ijk}$ is the Levi-Civitta tensor. Notice that summation over repeated 
indices is assumed.

As the derivative operator is a linear operator, only the extended dual 
version of the multiplication has to be obtained, but the addition has not.
Below we show the Fortran code for the extended dual version of the cross
product.

\begin{verbatim}
function crossdual(xd,yd) result(f_result)
	implicit none
	double precision :: f_result(3,3)
	double precision, intent(in) :: xd(3,3),yd(3,3)
	integer :: j,k
	
	f_result = 0d0
	
	do j=1,3
	 do k=1,3
	  f_result(1,:) = f_result(1,:) + levic(1,j,k)*proddual(xd(j,:),yd(k,:))
	  f_result(2,:) = f_result(2,:) + levic(2,j,k)*proddual(xd(j,:),yd(k,:))
	  f_result(3,:) = f_result(3,:) + levic(3,j,k)*proddual(xd(j,:),yd(k,:))
	 end do
	end do 

	return
end function crossdual
\end{verbatim}
In the {\tt levic} function, the Levi-Civita tensor is coded. In the {\tt proddual} function, 
the extended dual version of the scalar product is coded. The code for both 
functions is included in the downloadable module.
The result is a $3\times 3$ matrix where the first column is the real cross product, 
the second column is the first order derivative of the cross product, and the third 
column is the second order derivative of the cross product.

For example, let  ${\mathbf v}(\theta)=\left\{\cos (\theta ),\sin (\theta ),\theta 
^3\right\}$ and ${\mathbf w}(\theta)=\left\{e^{-\theta ^2},\theta  
\cos (\theta ),\sin (\theta )\right\}$ be two vectors. A program that calculates 
its extended dual cross product at $\theta = 1.1$ is as follows:
\begin{verbatim}
program crossp
use sphmodual
implicit none
double precision :: ang(3),v(3,3),w(3,3),thr(3),crsp(3,3)

thr = [3d0,0d0,0d0]
ang = [1.1d0,1d0,0d0]

v(1,:) = cosdual(ang)
v(2,:) = sindual(ang)
v(3,:) = powdual(ang,thr)

w(1,:) = expdual(-proddual(ang,ang))
w(2,:) = proddual(cosdual(ang),ang)
w(3,:) = sindual(ang)

crsp = crossdual(v,w)
print*,crsp(:,1)
print*,crsp(:,2)
print*,crsp(:,3)
end program crossp
\end{verbatim}

Extended dual version of other mathematical objects like dot product, norms, 
matrix multiplications, rotation matrices, etc., can be implemented in a similar way.

%% file: sec4.tex
\section{ First and second derivative in the spherical 4R mechanism}\label{sec4}
\begin{figure}[t]
\begin{center}
\includegraphics[scale=0.3]{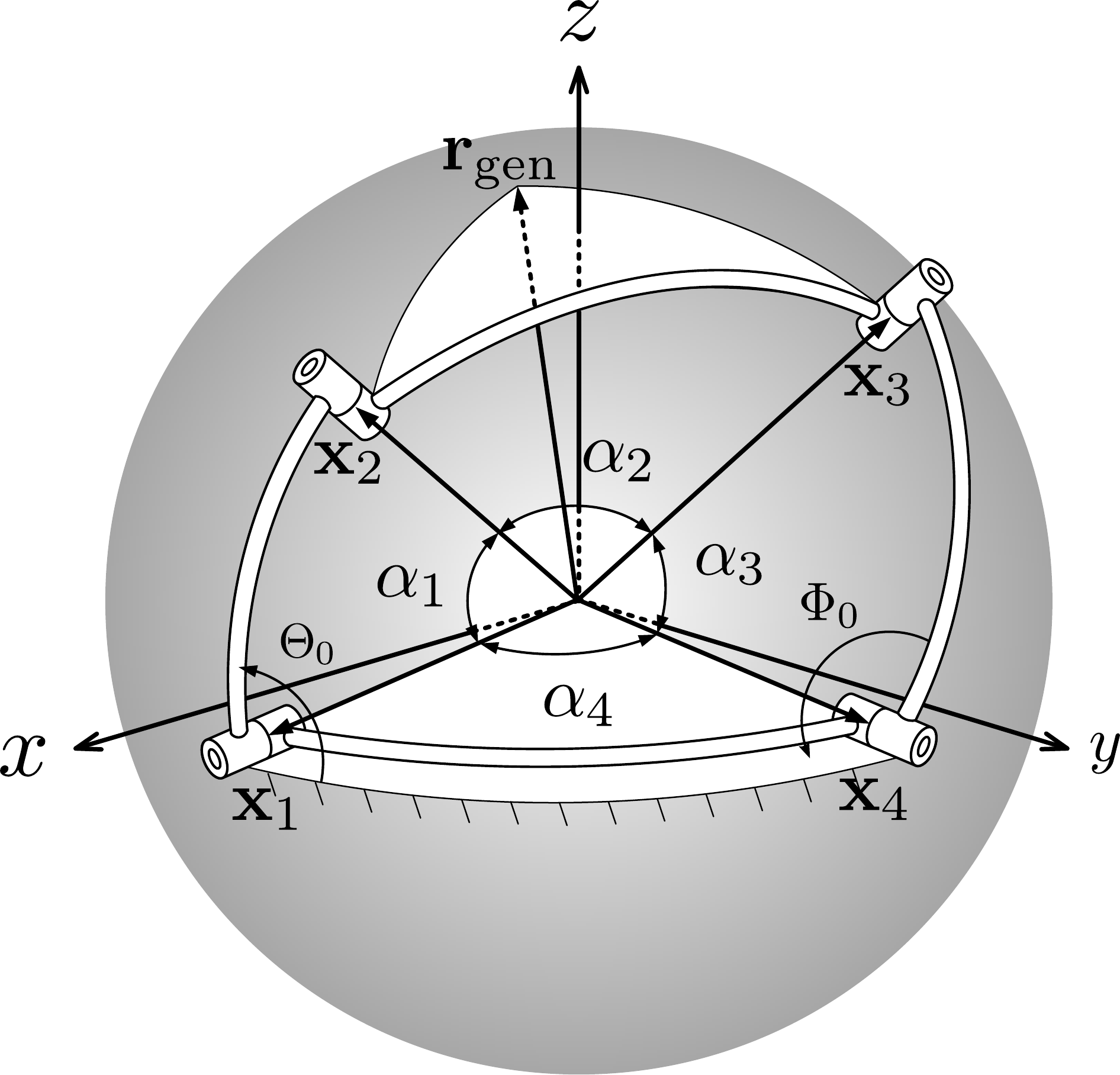} 
\caption{Spherical four-bar mechanism}
\label{figure1}
\end{center}
\end{figure}

This section presents the application of the proposed methodology to compute the 
first and second order derivatives of some useful functions in the synthesis of 
mechanisms. In particular, the examples concern the spherical four bar mechanism 
shown in Fig. \ref{figure1}.  A detailed procedure in order to obtain the coupler-point curve can be found in \cite{penunuri2012}. Here we reproduce the essential formulas in order to make the paper self-contained. Notice that the notation has been slightly changed in order to have a clear code for the  programming functions.

\subsection{Derivative of the output angle} 
 Let us consider the spherical four bar mechanism shown in Fig. \ref{figure1}.
Without loss of generality we assume a unit sphere. In its 
assembly configuration, the input angle is equal to $\Theta_0$ and the vectors 
${\mathbf x}_k$  $k= 1,2,3,4,$ represent the initial position vectors for the joints $k$, respectively. 
We will consider the input link as the geodesic connecting the points ${\mathbf x}_1$ 
and ${\mathbf x}_2$, the coupler link as the geodesic connecting the points 
${\mathbf x}_2$ and ${\mathbf x}_3$, the output link as the geodesic connecting the 
points ${\mathbf x}_3$ and ${\mathbf x}_4$. Finally the fixed link is the 
geodesic connecting ${\mathbf x}_4$ and ${\mathbf x}_1$. Since we are considering a unit sphere,  $\alpha_1$, $\alpha_2$, $\alpha_3$, $\alpha_4$
will be the lengths of the links, respectively.
Once the input link starts to rotate, for example by an angle $\theta$, its new 
position will be $\Theta_0 + \theta$. Similarly the position of the output link changes
by the amount $\phi(\theta;\vec{\mathbf{x}})$ where the symbol $\vec{\mathbf{x}}$ 
represents the dependence on ($\mathbf{x}_1$, $\mathbf{x}_2$, $\mathbf{x}_3$, 
$\mathbf{x}_4$).  In what follows the dependence 
of $\phi$ on $\vec{\mathbf{x}}$ will not be written to avoid unnecessary notation, 
however it will be written  in those vectors with such a dependence.

Let ${ \mathbf r}_2(\theta;\mathbf{x}_1, \mathbf{x}_2)$ and ${ \mathbf r}_3
(\phi(\theta);\vec{\mathbf{x}})$ denote the final positions of input and output link, respectively, given by 
\begin{align}\label{r2}
 {\mathbf {r}} _2(\theta;\mathbf{x}_1, \mathbf{x}_2)&={\mathbf R}(\theta,
 {\mathbf x}_1){\mathbf x}_2\\ \label{r3}
  {\mathbf r}_3(\phi(\theta);\vec{\mathbf{x}})&={\mathbf R}(\phi(\theta),
  {\mathbf x}_4){\mathbf x}_3,
\end{align}
where ${\mathbf R}(\vartheta,{\mathbf x})$ is a rotation matrix for 
angle $\vartheta$ about the unit vector ${\mathbf x}$.

Since the coupler link must have a constant length, the angle $ \phi (\theta)$ 
can be obtained from 

\begin{equation}\label{coplercondition}
{ \mathbf r}_2(\theta;\mathbf{x}_1, \mathbf{x}_2)\cdot {\mathbf r}_3(\phi(\theta);
\vec{\mathbf{x}})={\mathbf x}_2\cdot{\mathbf x}_3= \cos \alpha_2 =\text{constant}.
\end{equation}

A closed-form solution of Eq. (\ref{coplercondition}) can be found in 
\cite{Cervantes2009,Angeles2008}. In our notation, it is given by

\begin{align}\label{phiangleB}
\phi=\, \Phi_0  - 2 \tan^{-1}\left(\frac{A\pm \sqrt{A^2+B^2-C^2}}{C-B}\right),
\end{align}
where
\begin{align}
A=\,&\sin \alpha _1 \sin \alpha _3 \sin (\theta + \Theta_0) \nonumber\\
B=\,&\cos \alpha _1 \sin \alpha _3 \sin \alpha _4 -\sin \text{$\alpha _1$} 
\sin \text{$\alpha _3$} \cos
   \text{$\alpha _4$} \cos(\theta +  \Theta_0) \nonumber\\
 C=\,&\sin \text{$\alpha _1$} \cos \text{$\alpha _3 $} \sin \text{$\alpha _4$} 
 \cos(\theta + \Theta_0) +\cos \text{$\alpha _1$} \cos \text{$\alpha _3$}
   \cos \text{$\alpha _4$}-\cos \text{$\alpha _2 $}\nonumber.
\end{align}

Although we use Eq. (\ref{phiangleB}) to obtain the coupler-point curve generated 
by the mechanism, in this section, we are interested in showing a numerical 
approach, where the derivative of $\phi$ with respect to $\theta$ can be found 
without the analytical knowledge of the $\phi(\theta)$ angle. This can be obtained from $F(\theta,\phi)={ \mathbf r}_2(\theta)\cdot 
{\mathbf r}_3(\phi)$, as
\begin{equation} \label{dfdt1}
\frac{d\phi}{d\theta}(\theta) =-\frac{1}{\frac{\partial F}{\partial  \phi }
(\theta,\phi(\theta))} \, \frac{\partial F}{ \partial \theta}(\theta,\phi(\theta)).
\end{equation} 

In order to obtain a numerical value for Eq. (\ref{dfdt1}), the $\phi$ angle 
is obtained by numerically solving Eq. (\ref{coplercondition}).
Clearly the use of dual numbers is an advantage here, since by implementing the 
extended dual version of Eq. (\ref{coplercondition})  and applying the numerical 
method of solution also in the context of the extended dual functions, the first and second 
derivatives of Eq. (\ref{coplercondition}) with respect to $\theta$ are automatically 
obtained, along with the solution for $\phi$.
Notice that even when Eq. (\ref{dfdt1}) is a formal solution to the problem at hand, 
yet the derivatives still need to be calculated.

\subsection{Velocity and acceleration of the coupler point}
\begin{figure}[t]
\begin{center}
\includegraphics[scale=0.3]{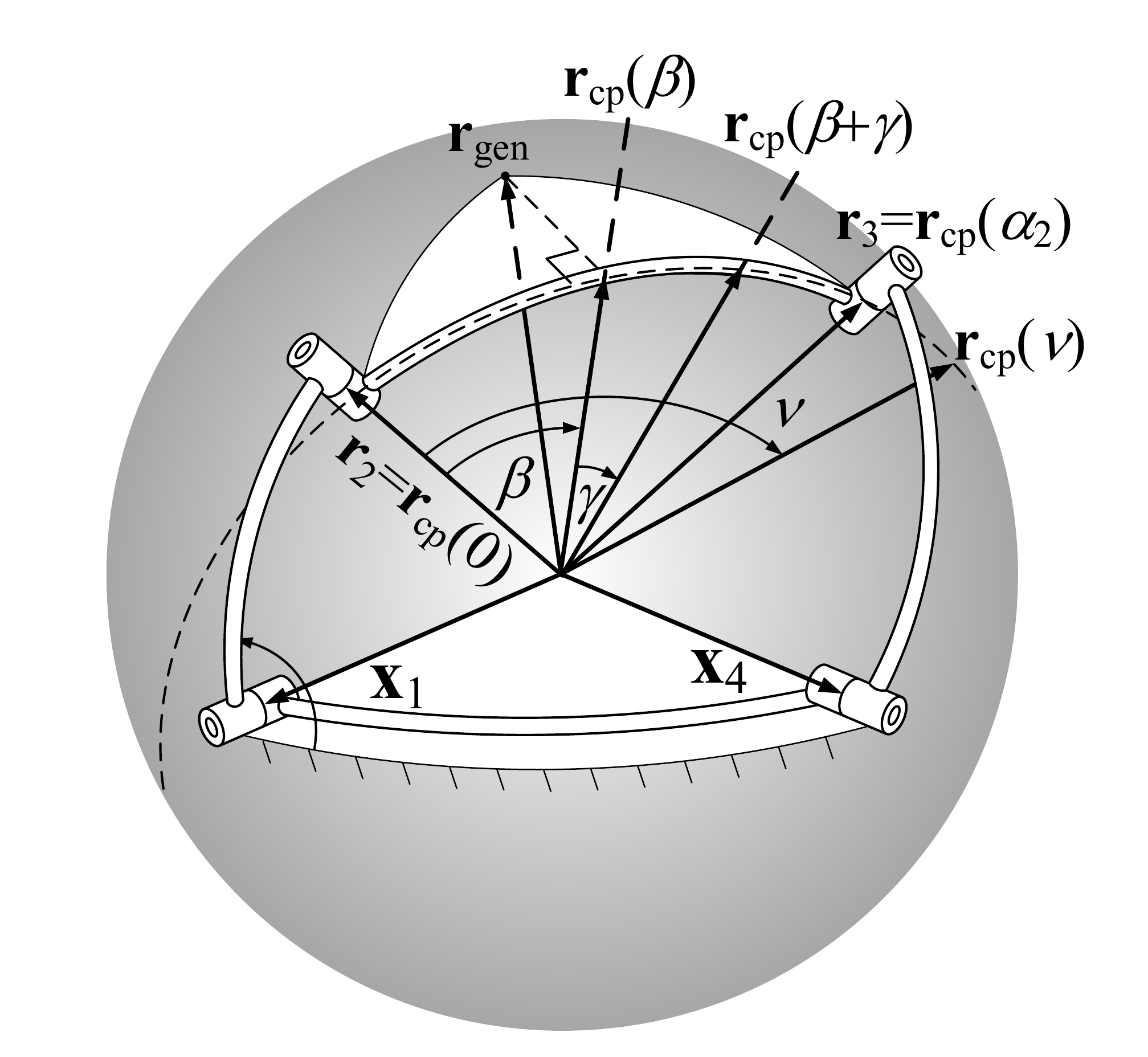} 
\caption{Vectors involved to obtain ${\mathbf r}_{\text{gen}}$.}
\label{figure2}
\end{center}
\end{figure}

We are interested in calculating the velocity and the acceleration of a point 
moving on the coupler-point curve.  Given a mechanism, the only independent 
(except time dependence) variable is the rotation angle $\theta$ of the input link. 
Denoting the position vector of the coupler point as $\mathbf{r}_{\text{gen}}$  
(see Fig. \ref{figure2}), we have
\begin{align}
\dot{\mathbf{r}}_\text{gen} &= \dot{\theta} \,\frac{\partial (\mathbf{r}_\text{gen})}
{\partial \theta} \\
\ddot{\mathbf{r}}_\text{gen} &= \dot{\theta}^2\, \frac{\partial^2 (\mathbf{r}_\text{gen})}
{\partial \theta^2} + \ddot{\theta} \, \frac{\partial (\mathbf{r}_\text{gen})}
{\partial \theta}, 
\end{align} 
for velocity and acceleration of $\mathbf{r}_\text{gen}$. 
 Since one can, in principle, control the angular velocity of the input 
 link, the problem is reduced to calculate $\partial (\mathbf{r}_\text{gen})/\partial 
 \theta$ and $\partial^2 (\mathbf{r}_\text{gen})/\partial \theta^2$.

In order to obtain a parametric equation for the coupler-point curve, it is necessary to find the position vector 
${\mathbf r}_{cp}(\theta,\nu;\vec{{\mathbf x}})$ of a point on the 
coupler link (see Fig. \ref{figure2}). This can be done by rotating the vector  
${\mathbf r}_2(\theta;{\mathbf x}_1,{\mathbf x}_2)$ by an angle $\nu$ about  
the unit vector  $\hat{{\mathbf n}}_{23}$ orthogonal to ${\mathbf r}_2
(\theta;{\mathbf x}_1,{\mathbf x}_2)$ and ${\mathbf r}_3(\phi(\theta);\vec{{\mathbf x}})$. 
Thus
\begin{equation}\label{couplerpoint}
\mathbf {r}_{cp}(\theta,\nu;\vec{{\mathbf x}})={\mathbf R}(\nu,\hat{{\mathbf n}}_{23})
{\mathbf r}_2(\theta;{\mathbf x}_1,{\mathbf x}_2)
\end{equation}
where 
\begin{equation}\label{n23}
\hat{\bf{n}}_{23}=\dfrac{{\mathbf r}_2(\theta;{\mathbf x}_1,{\mathbf x}_2) \times 
{\mathbf r}_3 (\phi(\theta);\vec{{\mathbf x}})} {\norm{{\mathbf r}_2(\theta;
{\mathbf x}_1,{\mathbf x}_2) \times {\mathbf r}_3(\phi(\theta);\vec{{\mathbf x}})}}. 
\end{equation}
Then, the position vector of the coupler point  
${\mathbf r}_{\text{gen}}$ is obtained by rotating the vector ${\mathbf r}_{cp}
(\theta,\beta + \gamma;\vec{{\mathbf x}})$, being $\gamma$  the angle between ${\mathbf r}_{cp}(\theta,\beta;\vec{{\mathbf x}})$ and $\mathbf{r}_{\text{gen}}$, by an angle of $\pi/2$ about
vector ${\mathbf r}_{cp}(\theta,\beta;\vec{{\mathbf x}})$:
\begin{equation}\label{rgenx}
{\mathbf r}_{\text{gen}}(\theta,\beta,\gamma;\vec{{\mathbf x}})={\mathbf R}
\left(\pi /2,{\mathbf r}_{cp}(\theta,\beta;\vec{{\mathbf x}}) \right)
{\mathbf r}_{cp}(\theta,\beta + \gamma;\vec{{\mathbf x}}).
\end{equation}

So, writing the Eq. (\ref{rgenx}) in its extended dual version, we could automatically 
get the parametric equation for the coupler-point curve, as well as its first and 
second derivatives with respect to $\theta$. 
It is worthwhile to mention that all the necessary functions to obtain  ${\mathbf r}_{\text{gen}}$ in its extended dual form are coded in the downloadable module.

As a practical example, let us consider the spherical four bar mechanism whose 
parameters are shown in Table \ref{pars4b}. This mechanism was synthesized for the path generation task  in \cite{penunuri2012}. The desired points were firstly presented in \cite{chusun2010} and normalized in \cite{Mullineux2011}. 

\begin{table}[htb]
\caption{Parameters of the Spherical 4R mechanism}
\centering
\scalebox{.85}{
\input{table1.tex}
}
\label{pars4b}
\end{table}

Table \ref{velocityacceleration} shows numerical results for velocity and 
acceleration,  of a point moving on the coupler-point curve as function 
of the input angle $\theta$. 

\begin{table}[htb]
\caption{Components of velocity vector $\dot{{\mathbf r}}_{\text{gen}}(\theta)$ and acceleration vector $\ddot{{\mathbf r}}_{\text{gen}}(\theta)$
for $\dot{\theta}=1 $ (units are in the SI system)}
\centering
\scalebox{.85}{
\input{table2.tex}
}
\label{velocityacceleration}
\end{table}

%% file: table1.tex
\begin{tabular}{cccccccccccc}
\toprule 
 \multicolumn{3}{c}{$\mathbf{x}_1$}&&\multicolumn{3}{c}{$\mathbf{x}_4$}  &$\alpha_1$
  & $\alpha_2$ &$\alpha_3$ & $\beta$ & $\gamma$ \\
 \cline{1-3} \cline{5-7} 
  1.00000 &0.00000 & 0.00000   &&  0.54462 & 0.80817 & 0.22413 & 0.40144 & 0.82034 & 0.92504 &  0.23067 & 0.47437\\
\bottomrule 
\end{tabular} 

%% file: table2.tex
\begin{tabular}{c r r r r r r}
\toprule 
 $\theta$& $\dot{x}_{\text{gen}}(\theta)$ &$\dot{y}_{\text{gen}}(\theta)$ &$\dot{z}_{\text{gen}}(\theta)$ &$\ddot{x}_{\text{gen}}(\theta)$ &$\ddot{y}_{\text{gen}}(\theta)$ &$\ddot{z}_{\text{gen}}(\theta)$ \\
  \midrule
 $0.00000$ & $-0.14255$ & $-0.06884$ & $0.59467$ & $ -0.42870$ &$ -0.25131$& $0.02053$\\
 $0.62832$ & $-0.28008$ & $-0.18548$ & $0.35545$ & $0.03897$ & $-0.15048$ & $-0.56498$\\
 $1.25664$ & $-0.17827$ & $-0.23972$ & $0.05446$ & $0.22578$ & $-0.00787$ & $-0.37389$\\
 $1.88496$ & $-0.02698$ & $-0.20190$ & $-0.13155$ & $0.24421$ & $0.11368$ & $-0.23910$\\
 $2.51327$ &  $0.10680$ & $-0.11271$ & $-0.26666$ & $0.15791$ & $0.16247$ & $-0.19681$\\
 $3.14159$ & $0.15218$ & $-0.00109$ & $-0.36590$ & $-0.00803$ & $0.19437$ & $-0.09747$\\
 $3.76991$ & $0.12511$ & $0.13203$ & $-0.36657$ & $-0.05418$ & $0.22196$ & $0.10407$\\
 $4.39823$ & $0.09935$ & $0.25462$ & $-0.23446$ & $-0.02558$ & $0.14294$ & $0.31101$\\
 $5.02655$ & $0.08944$ & $0.28061$ & $0.01394$ & $-0.01379$& $-0.08080 $ & $0.47129$\\
 $5.65487$ & $0.05510$ & $0.14226$ & $0.34717$ & $-0.14138$& $-0.34298$ & $0.56752$\\
   \bottomrule
\end{tabular}

%% file: conclusions.tex
\section{Conclusions}\label{sectionconclu}
 Velocity and acceleration of a  coupler-point on
 a spherical 4R mechanism are obtained by using  dual numbers. 
 Although it would be 
 possible to obtain an analytical expression for the position vector of the coupler point, 
 the complexity of the expression does not allow an efficient method to obtain its 
 derivatives, but, if the vector is written in its extended dual version, as it is 
 proposed in this work, its derivatives are obtained directly. As a consequence velocities and accelerations 
 can now be efficiently considered in the optimum synthesis of mechanisms.

The work details the proposal and implementation of a methodology to numerically 
obtain first and second order derivatives of one variable functions. With our proposal 
such derivatives (without any approximation) can be obtained in straightforward manner. 
The methodology is easy to implement and, although we have coded the extended dual 
functions in  the Fortran language, other programming languages could be used.

%% file: main.bbl
\begin{thebibliography}{10}
\expandafter\ifx\csname url\endcsname\relax
  \def\url#1{\texttt{#1}}\fi
\expandafter\ifx\csname urlprefix\endcsname\relax\def\urlprefix{URL }\fi
\expandafter\ifx\csname href\endcsname\relax
  \def\href#1#2{#2} \def\path#1{#1}\fi

\bibitem{Giacinto1981}
G.~Guj, Z.~Dong, M.~D. Giacinto, Dimensional synthesis of four bar linkage for
  function generation with velocity and acceleration constraints, Meccanica: an
  international journal of theoretical and applied mechanics 16~(4) (1981)
  210--219.

\bibitem{Russella2005}
K.~Russella, R.~S. Sodhib, On the design of slider-crank mechanisms. part {I}:
  multi-phase motion generation, Mechanism and Machine Theory 40~(3) (2005)
  285--299.

\bibitem{Sandgren1985}
E.~Sandgren, Design of single-and multiple-dwell six-link mechanisms through
  design optimization., Mechanism and Machine Theory 20~(6) (1985) 483--490.

\bibitem{Umit2007}
{\"Umit}.~{S\"onmez}, Introduction to compliant long dwell mechanism designs
  using buckling beams and arcs, Journal of Mechanical Design 129~(8) (2007)
  831 -- 844.

\bibitem{Mohan2009}
M.~Jagannath, S.~Bandyopadhyay, A new approach towards the synthesis of six-bar
  double dwell mechanisms, Computational Kinematics VII (2009) 209--216.

\bibitem{Zobairi2003}
M.~Zobairi, S.~Rao, B.~Sahay, Kineto-elastodynamic balancing of {4R-four} bar
  mechanisms combining kinematic and dynamic stress considerations, Mechanism
  and Machine Theory 21~(4) (1986) 307--315.

\bibitem{Jiang2007}
Y.-Q. Yu, B.~Jiang, Analytical and experimental study on the dynamic balancing
  of flexible mechanisms, Mechanism and Machine Theory 42~(5) (2007) 626--635.

\bibitem{Neidinger2010}
R.~D. Neidinger, Introduction to automatic differentiation and matlab
  object-oriented programming, SIAM Review 52~(3) (2010) 545--563.

\bibitem{Griewank1989}
A.~Griewank, On automatic differentiation, in Mathematical Programming: Recent
  Developments and Applications, M. Iri and K. Tanabe, eds., Klu wer Academic,
  Dordrecht, The Netherlands, 1998.

\bibitem{Study1901}
E.~Study, Geometrie der Dynamen, Verlag Teubner, Leipzig, 1903.

\bibitem{Clif1873}
W.~Clifford, Preliminary sketch of biquaternions, Proc. London Mathematical
  Society 1~(s1-4) (1873) 381--395.

\bibitem{Leuck1999}
H.~Leuck, H.-H. Nagel, Automatic differentiation facilitates of-integration
  into steering-angle-based road vehicle tracking, IEEE Computer Society
  Conference on Computer Vision and Pattern Recognition 2~(5) (1999) 2360.

\bibitem{Piponi2004}
D.~Piponi, Automatic differentiation, c++ templates, and photogrammetry,
  Journal of graphics, gpu, and game tools 9~(4) (2004) 41--55.

\bibitem{Jefrey2011}
J.~A. Fike, J.~J. Alonso (Eds.), The Development of Hyper-Dual Numbers for
  Exact Second-Derivative Calculations, Proceedings of the 49th AIAA Aerospace
  Sciences Meeting, Orlando Florida, USA, 2011.

\bibitem{Cheng1994}
H.~H. Cheng, Programming with dual numbers and its applications in mechanisms
  design, Engineering with Computers 10~(4) (1994) 212--229.

\bibitem{Ettore2008}
E.~Pennestr\`i, P.~Valentini, Linear dual algebra algorithms and their
  application to kinematics, Multibody Dynamics Computational Methods and
  Applications 12 (2008) 207--229.

\bibitem{Brodsky1999}
V.~Brodsky, M.~Shoham, Dual numbers representation of rigid body dinamics,
  Mechanism and Machine Theory 34 (1999) 975--991.

\bibitem{penunuri2012}
{F. Pe\~nu\~nuri}, {R. Pe\'on-Escalante}, C.~Villanueva, C.~A. Cruz-Villar,
  Synthesis of spherical 4r mechanism for path generation using differential
  evolution, Mechanism and Machine Theory 57 (2012) 62--70.

\bibitem{Cervantes2009}
J.~J. Cervantes-S\'anchez, H.~I. Medell\'in-Castillo, J.~M. Rico-Mart\'inez,
  E.~J. Gonz\'alez-Galv\'an, Some improvements on the exact kinematic synthesis
  of spherical 4r function generators, Mechanism and Machine Theory 44 (2009)
  103--121.

\bibitem{Angeles2008}
S.~Bai, J.~Angeles, A unified input-output analysis of four-bar linkages,
  Mechanism and Machine Theory 43 (2008) 240--251.

\bibitem{chusun2010}
J.~Chu, J.~Sun, Numerical atlas method for path generation of spherical
  four-bar mechanism, Mechanism and Machine Theory 45 (2010) 867--879.

\bibitem{Mullineux2011}
G.~Mullineux, Atlas of spherical four-bar mechanisms, Mechanism and Machine
  Theory 46 (2011) 1811--1823.

\end{thebibliography}
